\documentclass[aps,prd,preprint,onecolumn,nofootinbib,superscriptaddress]{revtex4-2}

\usepackage[utf8]{inputenc}
\usepackage[T1]{fontenc}
\usepackage{lmodern}

\usepackage{amsmath,amssymb,amsfonts,mathtools}
\usepackage{bm}
\usepackage{graphicx}
\usepackage{microtype}
\usepackage{siunitx}
\usepackage{xspace}
\usepackage{booktabs}
\usepackage{hyperref}

\hypersetup{
  colorlinks=true,
  linkcolor=blue,
  citecolor=blue,
  urlcolor=blue,
  pdfauthor={Jorge Lamas},
  pdftitle={Emergent spectral geometry in the Coherence--Curvature Model}
}

\graphicspath{{figs/}}

\newcommand{\CCM}{\textsc{CCM}\xspace}
\newcommand{\ds}{d_s}
\newcommand{\dhh}{d_h}
\newcommand{\mean}[1]{\langle #1 \rangle}

\newcommand{\abs}[1]{\left\lvert #1 \right\rvert}

\newcommand{\diff}{\mathrm{d}}

\newcommand{\Graph}{G}
\newcommand{\Vertices}{V}
\newcommand{\Edges}{E}
\newcommand{\Laplace}{L}

\newcommand{\Ric}{R_{O}}

\begin{document}

\newcommand{\NmaxFSS}{1024}
\newcommand{\DsNmaxValue}{3.702}
\newcommand{\DsNmaxError}{1.152}
\newcommand{\DhNmaxValue}{2.888}
\newcommand{\DhNmaxError}{0.044}
\newcommand{\DsOverDhNmaxValue}{1.282}
\newcommand{\DsOverDhNmaxError}{0.399}
\newcommand{\DistScalingExponent}{0.126}

\title{Emergent spectral geometry in the Coherence--Curvature Model}

\author{Jorge Lamas}
\email{j0rglms@gmail.com}
\affiliation{Independent researcher}

\date{\today}

\begin{abstract}
We investigate the Coherence--Curvature Model (\CCM), a dynamical ensemble of connected graphs governed by a Hamiltonian that couples global spectral coherence to local Ollivier--Ricci curvature and to a tunable edge-density penalty. Using connected simulated annealing, we generate low-energy configurations of simple, undirected graphs and characterize their emergent geometry through the spectral dimension $\ds$, the Hausdorff dimension $\dhh$, and the average graph distance $\mean{d}$. Finite-size scaling analyses based on ensembles with up to $N=\NmaxFSS$ vertices show a clear growth of the spectral dimension with system size, accompanied by a slower increase of the Hausdorff dimension. At the largest simulated volume we estimate
\[
  d_s = \DsNmaxValue \pm \DsNmaxError,\qquad
  d_h = \DhNmaxValue \pm \DhNmaxError,\qquad
  \frac{d_s}{d_h} = \DsOverDhNmaxValue \pm \DsOverDhNmaxError.
\]
Within the current uncertainties, these values indicate a clear hierarchy where $\ds > \dhh$, with the spectral dimension potentially approaching four, while the Hausdorff dimension robustly stabilizes around a value close to three. Crucially, we demonstrate that these dimensions are decoupled: the spectral dimension can be independently tuned by the curvature coupling $\gamma$, while the Hausdorff dimension remains comparatively rigid. We further study the dependence on the locality coupling $\beta$, and we analyze the scaling of the mean graph distance $\mean{d}(N)$, which follows a slow power law $\mean{d}(N)\sim N^\eta$ with $\eta = \DistScalingExponent$, indicative of a small-world regime. Overall, the \CCM\ provides a controlled numerical laboratory in which the interplay between spectral coherence, curvature, and edge density gives rise to finite-dimensional, fractal-like geometries with tunable, scale-dependent effective dimension.
\end{abstract}

\maketitle

\section{Introduction}
\label{sec:introduction}

Discrete approaches to quantum geometry and spacetime seek to understand how continuum-like structures can emerge from fundamentally combinatorial degrees of freedom. Examples include causal dynamical triangulations, group field theories, spin foam models, random geometric graphs, and a variety of network-based formulations of emergent geometry and quantum gravity~\cite{Ambjorn:2012CDT,Oriti:2011GFT}. A central question common to these frameworks is whether ensembles of discrete configurations can exhibit effective large-scale properties that resemble those of smooth manifolds, and if so, how these properties depend on microscopic parameters and dynamics.

Two key geometric observables in this context are the \emph{spectral dimension} $\ds$ and the \emph{Hausdorff dimension} $\dhh$. The spectral dimension is defined via diffusion processes on the underlying graph and probes how a random walker explores the configuration as a function of diffusion time. The Hausdorff dimension, by contrast, is obtained from how volumes of metric balls scale with their radius. Together, $\ds$ and $\dhh$ provide complementary information about spectral and volumetric aspects of the emergent geometry, and discrepancies between them can signal nontrivial scale dependence, fractality, or multifractal behavior~\cite{Ambjorn:2012CDT,Bollobas:2001RandomGraphs,Barabasi:2016book}.

Curvature is another fundamental geometric quantity, and several notions of discrete curvature have been proposed for graphs and networks. Among these, Ollivier--Ricci curvature, defined in terms of optimal transport between local probability measures, has attracted significant attention as a flexible and computationally accessible notion of curvature on general metric spaces and graphs~\cite{Ollivier:2009Ricci}. Ollivier--Ricci curvature has been used to analyze complex networks, study robustness and community structure~\cite{Ni:2019Ricci}, and explore geometric features of discrete models, with efficient implementations based on entropic-regularized optimal transport and Sinkhorn iterations~\cite{Cuturi:2013Sinkhorn,Flamary:2021POT}.

In this work we introduce and analyze the Coherence--Curvature Model (\CCM), a dynamical graph model in which the Hamiltonian balances a global spectral term, built from the algebraic connectivity (the second eigenvalue of the graph Laplacian), against a local curvature term based on Ollivier--Ricci curvature, together with a simple contribution proportional to the number of edges. In contrast to previous studies of fixed-degree ensembles, we consider here simple, undirected, \emph{connected} graphs with a variable number of edges. The edge-count term in the Hamiltonian is therefore genuinely dynamical and acts as a locality coupling: increasing its coefficient penalizes dense graphs and favors sparser, more locally structured configurations.

The \CCM\ is explored numerically by connected simulated annealing, a version of simulated annealing in which proposed graph updates are constrained to keep the graph connected at all times. For each set of parameters we generate ensembles of low-energy configurations and measure the spectral dimension $\ds$, the Hausdorff dimension $\dhh$, and the average graph distance $\mean{d}$ as functions of the system size and of the couplings. The numerical results, summarized by the macros in \texttt{data/results\_summary.tex}, indicate an emergent spectral geometry consistent with a finite effective dimension and a clear hierarchy $\ds > \dhh$ at the largest volumes, together with a slow growth of typical distances with $N$.

The structure of the paper is as follows. In Sec.~\ref{sec:model-methods} we define the \CCM\ Hamiltonian, introduce the geometric observables, and describe the connected-simulated-annealing algorithm used to sample the ensemble. Section~\ref{sec:results} presents the numerical results, including finite-size scaling, convergence with annealing depth, and the dependence on both the curvature coupling and the locality coupling. In Sec.~\ref{sec:discussion} we interpret these findings in the broader context of emergent geometry and discuss their implications for discrete gravity-inspired models. Finally, Sec.~\ref{sec:conclusion} summarizes our main conclusions and outlines directions for future work.

\section{Model and Methods}
\label{sec:model-methods}

In this section we provide a precise definition of the Coherence--Curvature Model and of the observables used to characterize its emergent geometry. We also summarize the numerical strategy employed to sample low-energy configurations and to estimate effective dimensions.

\subsection{Coherence--Curvature Hamiltonian}
\label{subsec:hamiltonian}

The configurations of the \CCM\ are simple, undirected, connected graphs
\begin{equation}
  \Graph = (\Vertices,\Edges),
\end{equation}
with $\Vertices$ the set of $N = \abs{\Vertices}$ labeled vertices and $\Edges$ the set of edges. The number of edges $\abs{\Edges}$ is dynamical and will be controlled by one of the couplings in the Hamiltonian. Connectivity is imposed as a hard constraint: only connected graphs contribute to the ensemble, and all update moves preserve connectivity.

To each graph we associate the combinatorial Laplacian
\begin{equation}
  \Laplace = D - A,
\end{equation}
where $A$ is the adjacency matrix and $D$ is the diagonal degree matrix, $D_{vv} = \deg(v)$. The eigenvalues of $\Laplace$ are denoted
\begin{equation}
  0 = \lambda_1(\Graph) \leq \lambda_2(\Graph) \leq \cdots \leq \lambda_N(\Graph).
\end{equation}
The quantity $\lambda_2(\Graph)$ is the \emph{algebraic connectivity} or Fiedler value, which measures global connectivity and coherence of the graph.

The energy of a configuration $\Graph$ in the \CCM\ is defined by the Hamiltonian
\begin{equation}
  H(\Graph)
  =
  -\alpha\,\lambda_2(\Graph)
  + \beta\,\abs{\Edges}
  + \gamma \sum_{v \in \Vertices} \Ric(v),
  \label{eq:hamiltonian}
\end{equation}
where $\alpha$, $\beta$, and $\gamma$ are real parameters, and $\Ric(v)$ is the Ollivier--Ricci curvature associated with vertex $v$, defined as an average over the curvature of incident edges.

The three terms in Eq.~\eqref{eq:hamiltonian} play distinct roles:
\begin{itemize}
  \item The spectral coherence term $-\alpha\,\lambda_2(\Graph)$ favors graphs with large algebraic connectivity. For fixed $N$, such graphs tend to have short typical distances and strong global connectivity.
  \item The edge-count term $+\beta\,\abs{\Edges}$ penalizes dense graphs. For $\beta>0$ this contribution suppresses the proliferation of edges and promotes sparsity and locality, counteracting the tendency of the spectral term to produce overly dense graphs.
  \item The curvature term $\gamma \sum_{v} \Ric(v)$ couples to the average Ollivier--Ricci curvature. Since this curvature is typically negative on tree-like structures and can become less negative or positive in the presence of local clustering and short cycles, the sign and magnitude of $\gamma$ control the model's preference for different curvature regimes.
\end{itemize}
In the main parameter scans we fix $(\alpha,\beta,\gamma)=(1.0,0.0091,0.15)$ and vary the graph size $N$, the annealing depth, and the curvature coupling $\gamma$. In a separate locality scan we keep $\alpha$ and $\gamma$ fixed and vary $\beta$ to explore how the edge-density penalty affects the emergent geometry.

\subsection{Ollivier--Ricci curvature}
\label{subsec:orc}

We briefly recall the definition of Ollivier--Ricci curvature for graphs and the specific implementation used in the simulations~\cite{Ollivier:2009Ricci,Ni:2019Ricci}.

Given a connected graph $\Graph=(\Vertices,\Edges)$ with graph distance $d(\cdot,\cdot)$, we associate to each vertex $v\in\Vertices$ a probability measure $\mu_v$ supported on $v$ and its neighbors. In our implementation we use a lazy random walk with idleness parameter $\alpha_{\mathrm{ORC}}=0.5$: with probability $\alpha_{\mathrm{ORC}}$ the walker stays at $v$, and with probability $1-\alpha_{\mathrm{ORC}}$ it moves to one of the neighbors chosen uniformly at random. The measure $\mu_v$ is thus
\begin{equation}
  \mu_v(u)
  =
  \begin{cases}
    \alpha_{\mathrm{ORC}}, & u = v, \\
    \dfrac{1-\alpha_{\mathrm{ORC}}}{\deg(v)}, & u \sim v, \\
    0, & \text{otherwise},
  \end{cases}
\end{equation}
where $u \sim v$ denotes adjacency.

For two adjacent vertices $u$ and $v$ we define the Ollivier--Ricci curvature $\kappa_{uv}$ by
\begin{equation}
  \kappa_{uv}
  =
  1 - \frac{W_1(\mu_u,\mu_v)}{d(u,v)},
\end{equation}
where $W_1$ is the 1-Wasserstein (Earth mover's) distance between probability measures on $\Vertices$ with respect to the graph distance $d$~\cite{Ollivier:2009Ricci}. On unweighted graphs, $d(u,v)=1$ for neighbors, so the denominator simply normalizes the transport distance. The vertex-wise curvature $\Ric(v)$ appearing in Eq.~\eqref{eq:hamiltonian} is taken as the average of the edge curvatures incident on $v$,
\begin{equation}
  \Ric(v) = \frac{1}{\deg(v)} \sum_{u \sim v} \kappa_{uv}.
\end{equation}

The optimal transport problem defining $W_1(\mu_u,\mu_v)$ is solved numerically using an entropic-regularized scheme based on the Sinkhorn algorithm~\cite{Cuturi:2013Sinkhorn}, as implemented in the \textsc{GraphRicciCurvature} Python library~\cite{Ni:2019Ricci}, which in turn relies on the \textsc{POT} (Python Optimal Transport) toolbox~\cite{Flamary:2021POT}. The entropic regularization provides a good compromise between numerical stability and computational cost for the graph sizes considered here.

\subsection{Geometric observables}
\label{subsec:observables}

The emergent geometry of the \CCM\ ensembles is characterized by three main observables: the spectral dimension $\ds$, the Hausdorff dimension $\dhh$, and the average graph distance $\mean{d}$. We summarize their operational definitions and the fitting procedures used in the analysis.

\subsubsection{Spectral dimension}

The spectral dimension $\ds$ is defined in terms of a discrete diffusion process on the graph. Consider a simple random walk that starts at a vertex $i$ at time $t=0$ and moves at each step to a uniformly chosen neighbor. Let $P_{ii}(t)$ denote the probability that the walker is back at $i$ after $t$ steps. On a homogeneous infinite space the average return probability $\mean{P(t)}$ behaves as
\begin{equation}
  \mean{P(t)} \sim t^{-\ds/2}
  \label{eq:spectral-dim-ansatz}
\end{equation}
for an appropriate range of diffusion times. This motivates the definition of an effective, time-dependent spectral dimension
\begin{equation}
  d_s(t)
  =
  -2\,\frac{\diff\ln \mean{P(t)}}{\diff\ln t}.
  \label{eq:ds-effective}
\end{equation}
On a finite graph, $d_s(t)$ typically shows a plateau at intermediate times, from which a scale-dependent estimate of $\ds$ can be extracted.

In practice, for each graph $\Graph$ we perform a large number of independent random walks (in our simulations, 2000 walks per configuration), with starting vertices chosen uniformly at random. For each walk we record the return probability up to a maximum diffusion time $t_{\max}$, and we then average over walks and over configurations. The effective dimension $d_s(t)$ is obtained from a discrete logarithmic derivative of $\mean{P(t)}$, and a single value of $\ds$ is extracted by fitting Eq.~\eqref{eq:spectral-dim-ansatz} in a window of diffusion times $t \in [t_{\min}, t_{\max}]$ where $d_s(t)$ is approximately flat. In the parameter choices used here, we typically take $t_{\min}=2$ and $t_{\max}=16$, with the precise fit window chosen to maximize the stability of the fit and minimize systematic drift.

\subsubsection{Hausdorff dimension}

The Hausdorff dimension $\dhh$ is obtained from the scaling of ball volumes with radius. For each configuration we select a set of randomly chosen center vertices $i$ (32 centers in the simulations reported here). For each center we define balls
\begin{equation}
  B(i,r) = \{ j \in \Vertices \,\vert\, d(i,j) \leq r \},
\end{equation}
where $d(i,j)$ is the geodesic (shortest-path) distance. The volume $V(i,r)$ of the ball is simply the cardinality $\abs{B(i,r)}$, and we define the averaged volume
\begin{equation}
  V(r) = \mean{V(i,r)}_{i,\text{configurations}},
\end{equation}
where the average is taken over centers and over Monte Carlo configurations. On a homogeneous infinite space one expects the scaling law
\begin{equation}
  V(r) \sim r^{\dhh}
  \label{eq:volume-scaling}
\end{equation}
for intermediate radii. On finite graphs there are small-$r$ corrections and a saturation at large $r$ due to the finite diameter. We therefore determine $\dhh$ from a log--log fit of $\ln V(r)$ versus $\ln r$ in an intermediate window $r \in [r_{\min},r_{\max}]$, with $r_{\min}=1$ and $r_{\max}$ chosen such that $r_{\max}$ remains well below the graph diameter while still sampling a nontrivial range of scales. The quoted uncertainties include both statistical errors from the fit and an estimate of the systematic variation under modest shifts of the fit window.

\subsubsection{Average distance and finite-size scaling}

The third observable is the average graph distance
\begin{equation}
  \mean{d}
  =
  \frac{1}{N(N-1)} \sum_{i\neq j} d(i,j),
\end{equation}
which probes typical path lengths between vertices. For each configuration we compute all-pairs shortest-path distances using standard algorithms and average over configurations and seeds.

To characterize the finite-size scaling of $\mean{d}$ we assume a power-law ansatz
\begin{equation}
  \mean{d}(N) \sim N^\eta,
  \label{eq:distance-scaling}
\end{equation}
for some effective exponent $\eta$. This form is not intended as a fundamental law; on the range of sizes available ($N=128$--$\NmaxFSS$) it provides a convenient one-parameter characterization of the observed slow growth of distances. The exponent $\eta$ is obtained from a fit of $\ln \mean{d}(N)$ versus $\ln N$, and for the main parameter choices its numerical value is
\begin{equation}
  \eta = \DistScalingExponent.
\end{equation}

\subsection{Connected simulated annealing}
\label{subsec:numerics}

We explore the configuration space of the \CCM\ using a connected simulated-annealing (CSA) algorithm. The state space consists of simple, undirected, connected graphs on $N$ labeled vertices, with a variable number of edges. Connectivity is enforced at all times.

\paragraph*{Initialization.}
For each run we initialize the graph as a uniformly random labeled tree on $N$ vertices, generated for instance by a Prüfer sequence. This provides a connected, minimally dense starting configuration with $\abs{\Edges}=N-1$.

\paragraph*{Monte Carlo moves.}
One Monte Carlo step is defined as a single proposed graph update of one of two types:
\begin{enumerate}
  \item \emph{Edge addition:} select two distinct vertices $u$ and $v$ that are not currently adjacent, and propose to add the edge $(u,v)$.
  \item \emph{Edge deletion:} select an existing edge $(u,v)\in\Edges$ uniformly at random and propose to delete it, \emph{subject to} the constraint that the resulting graph remains connected.
\end{enumerate}
In practice, we alternate between addition and deletion attempts with equal probability. For deletion moves we test connectivity (e.g., via a breadth-first search) after removing the edge; if the graph becomes disconnected, the move is rejected and the previous graph is restored without evaluating the energy. Proposals that preserve connectivity are accepted or rejected according to the Metropolis rule described below.

\paragraph*{Annealing schedule and acceptance rule.}
We use a temperature schedule of the form
\begin{equation}
  T(k)
  =
  \frac{T_0}{\bigl(1 + k/k_0\bigr)^{\nu}},
\end{equation}
where $k$ is the Monte Carlo step index, $T_0$ is the initial temperature, $k_0$ sets the crossover scale, and $\nu$ controls the cooling rate. In the simulations reported here we use
\begin{equation}
  T_0 = 1.0,\qquad k_0 = 1000,\qquad \nu = 1.0.
\end{equation}
Given a proposed move $\Graph \to \Graph'$, the energy difference $\Delta H = H(\Graph') - H(\Graph)$ is computed using Eq.~\eqref{eq:hamiltonian}, with local updates wherever possible. The move is accepted with Metropolis probability
\begin{equation}
  p_{\mathrm{acc}} = \min\bigl(1, e^{-\Delta H / T(k)}\bigr).
\end{equation}
Each simulation consists of $S_{\max}$ Monte Carlo steps starting from the initial tree. We then discard an initial fraction of the trajectory as burn-in and measure observables on the remaining portion, averaging over a set of independent runs with different random seeds.

\paragraph*{Parameter sets and statistics.}
The main simulation campaigns are:
\begin{itemize}
  \item \emph{Finite-size scaling (FSS):} $N \in \{128,256,512,\NmaxFSS\}$, fixed couplings $(\alpha,\beta,\gamma)=(1.0,0.0091,0.15)$, and $S_{\max}=16000$ steps per run.
  \item \emph{Convergence (CONV):} $N=512$, $(\alpha,\beta,\gamma)=(1.0,0.0091,0.15)$, and $S_{\max}\in\{4000,16000,32000\}$, to assess the dependence on the annealing depth.
  \item \emph{Curvature sweep (GAMMA):} $N=256$, $(\alpha,\beta,\gamma)=(1.0,0.0091,\gamma)$ with $\gamma\in\{0.05,0.15,0.30,0.60,1.20\}$, and $S_{\max}=16000$.
  \item \emph{Locality sweep (BETA):} $N=256$, $(\alpha,\beta,\gamma)=(1.0,\beta,0.15)$ with $\beta\in\{0.0050,0.0091,0.0150,0.0250,0.0400\}$, and $S_{\max}=16000$.
\end{itemize}
For each parameter set we generate 10 independent runs (seeds). The main simulation parameters are summarized in Table~\ref{tab:sim-params}.

\begin{table}[t]
  \centering
  \caption{Summary of simulation parameters used in this work. Here $S_{\max}$ is the maximum number of Monte Carlo steps per run, and ``seeds'' is the number of independent runs per parameter set. The Ollivier--Ricci curvature is computed with idleness $\alpha_{\mathrm{ORC}}=0.5$ using entropic-regularized optimal transport (Sinkhorn algorithm).}
  \label{tab:sim-params}
  \begin{tabular}{llllll}
    \toprule
    Study & $N$ & $S_{\max}$ & $(\alpha,\beta,\gamma)$ & seeds & purpose \\
    \midrule
    FSS   & $128,256,512,\NmaxFSS$ & $16000$ & $(1.0,0.0091,0.15)$ & $10$ & finite-size scaling \\
    CONV  & $512$                  & $4000,16000,32000$ & $(1.0,0.0091,0.15)$ & $10$ & annealing depth \\
    GAMMA & $256$                  & $16000$ & $(1.0,0.0091,\gamma)$, $\gamma\in\{0.05,\dots,1.20\}$ & $10$ & curvature coupling \\
    BETA  & $256$                  & $16000$ & $(1.0,\beta,0.15)$, $\beta\in\{0.0050,\dots,0.0400\}$ & $10$ & locality coupling \\
    \bottomrule
  \end{tabular}
\end{table}

For each ensemble we measure the energy density, algebraic connectivity $\lambda_2$, mean Ollivier--Ricci curvature, spectral and Hausdorff dimensions, and average distance. Aggregated statistics are stored in CSV files in \texttt{paper/data/} and summarized for publication by the macros in \texttt{data/results\_summary.tex}.

\section{Results}
\label{sec:results}

In this section we present the numerical results obtained from the connected simulated-annealing exploration of the \CCM\ ensemble. Unless otherwise specified, quoted uncertainties correspond to one standard deviation across seeds, combined with the fitting uncertainties where relevant.

\subsection{Finite-size scaling at fixed curvature coupling}
\label{subsec:results-fss}

\begin{figure}[t]
  \centering
  \includegraphics[width=0.7\linewidth]{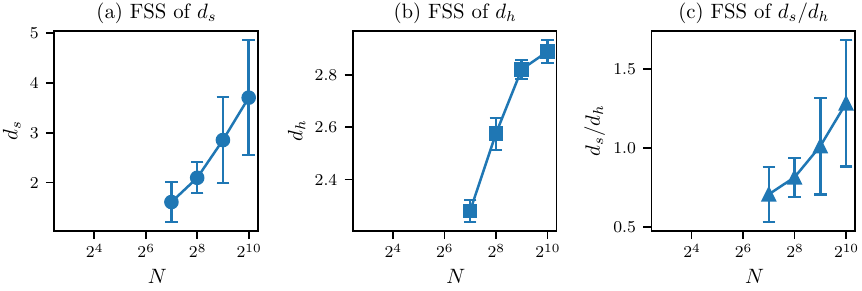}
  \caption{Finite-size scaling of the \CCM\ ensemble at fixed couplings $(\alpha,\beta,\gamma) = (1.0, 0.0091, 0.15)$ and annealing depth $S_{\max} = 16000$ steps. The panels show (a) the spectral dimension $d_s$, (b) the Hausdorff dimension $d_h$, and (c) the ratio $d_s/d_h$ as functions of the number of vertices $N \in \{128,256,512,\NmaxFSS\}$. Error bars denote one standard deviation across seeds, including fitting uncertainties where applicable. The data indicate a growth of $d_s$ with $N$ towards values compatible with four within current errors, a milder increase of $d_h$ towards values close to three, and a slow growth of typical distances consistent with the scaling form in Eq.~\eqref{eq:distance-scaling}.}
  \label{fig:FSS}
\end{figure}

We begin with the finite-size scaling study at fixed couplings $(\alpha,\beta,\gamma)=(1.0,0.0091,0.15)$ and $S_{\max}=16000$ steps, for $N\in\{128,256,512,\NmaxFSS\}$. The main observables are shown in Fig.~\ref{fig:FSS}, which displays the spectral dimension $d_s$, the Hausdorff dimension $d_h$, and the ratio $d_s/d_h$ as functions of $N$. The average distance $\mean{d}$, although not shown in the figure, exhibits a similarly smooth finite-size dependence, discussed below.

At the smallest volume, $N=128$, the spectral dimension is significantly below four and the estimates exhibit sizable statistical fluctuations due to the limited scale separation between the microscopic and macroscopic regimes. As $N$ increases, $\ds$ grows monotonically within errors, reaching $d_s = \DsNmaxValue \pm \DsNmaxError$ at the largest simulated size $N=\NmaxFSS$. The large uncertainty (approximately 31\% of the mean) reflects both the slow convergence of spectral properties (see Sec.~\ref{subsec:results-conv}) and the statistical fluctuations inherent in the measurement. While the central value is close to four, the confidence interval spans a wide range, making it difficult to statistically distinguish between $\ds=3$ and $\ds=4$ based on the current data. The increase of $\ds$ with $N$ suggests that the spectral dimension is still subject to finite-size effects and that the effective large-scale diffusion properties continue to evolve towards a higher-dimensional regime as the system size grows.

In contrast, the Hausdorff dimension $\dhh$ increases more modestly with $N$. For the smallest graphs it is already of order two, and it gradually approaches $d_h = \DhNmaxValue \pm \DhNmaxError$ at $N=\NmaxFSS$. Within the present uncertainties this is consistent with a volumetric dimension close to three, significantly below the putative spectral dimension but clearly larger than two. The ratio
\begin{equation}
  \frac{d_s}{d_h}
  = \DsOverDhNmaxValue \pm \DsOverDhNmaxError,
\end{equation}
thus quantifies a nontrivial hierarchy between spectral and volumetric notions of dimension in the \CCM\ ensemble.

The average distance $\mean{d}$ shows a slow but steady increase with $N$, from values slightly above three at $N=128$ to values slightly above four at $N=\NmaxFSS$. A fit to the scaling form \eqref{eq:distance-scaling} yields an effective exponent $\eta = \DistScalingExponent$, indicating that typical distances grow much more slowly than any power law with exponent of order one. This behavior is compatible with a small-world or near-small-world regime over the explored range of sizes, albeit with an effective volumetric dimension that remains clearly finite.

\subsection{Annealing depth and convergence at $N=512$}
\label{subsec:results-conv}

\begin{figure}[t]
  \centering
  \includegraphics[width=0.7\linewidth]{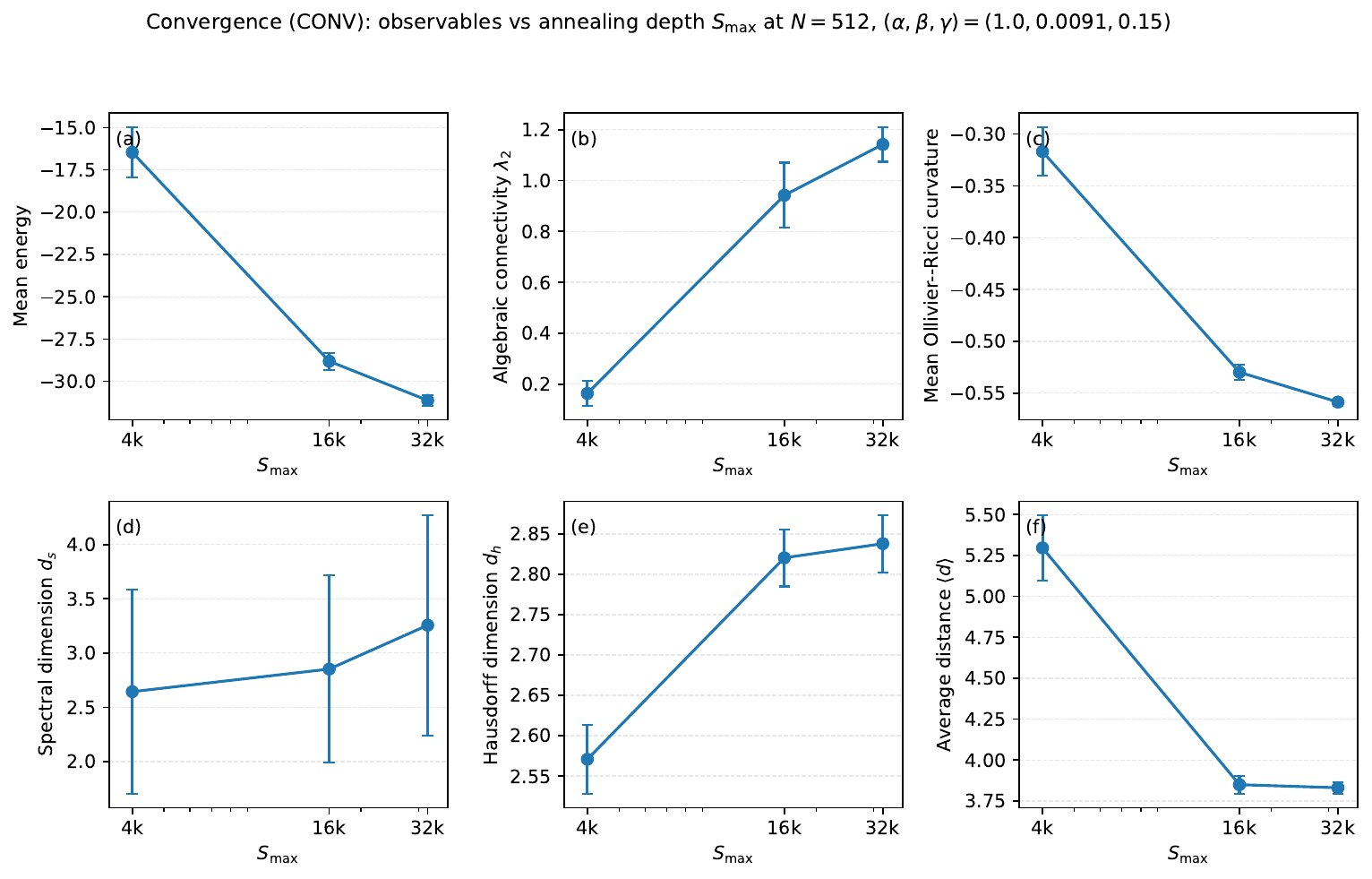}
  \caption{Annealing-depth study at fixed size $N=512$ and couplings $(\alpha,\beta,\gamma)=(1.0,0.0091,0.15)$. The panels show the dependence of energy density, algebraic connectivity $\lambda_2$, mean Ollivier--Ricci curvature, spectral dimension $d_s$, Hausdorff dimension $d_h$, and average distance $\mean{d}$ on the maximum number of Monte Carlo steps $S_{\max}\in\{4000,16000,32000\}$. Error bars denote one standard deviation across seeds. The results indicate a slow convergence of the spectral properties, with $d_s$ increasing for deeper annealing, while $d_h$ and $\mean{d}$ exhibit milder variations and stabilize more rapidly.}
  \label{fig:Convergence}
\end{figure}

To assess the impact of annealing depth and the approach to equilibrium, we perform convergence tests at fixed size $N=512$ and couplings $(\alpha,\beta,\gamma)=(1.0,0.0091,0.15)$, varying the maximum number of steps $S_{\max}\in\{4000,16000,32000\}$. The corresponding observables are summarized in Fig.~\ref{fig:Convergence}.

As expected, the energy density decreases as $S_{\max}$ increases, and the algebraic connectivity $\lambda_2$ grows, indicating that the longer runs explore configurations with higher spectral coherence. The mean Ollivier--Ricci curvature becomes more negative with increasing $S_{\max}$, consistent with the graphs developing locally more tree-like or hyperbolic features as the annealing proceeds.

The geometric observables also exhibit a dependence on the annealing depth. The spectral dimension $\ds$ increases from relatively low values at $S_{\max}=4000$ to larger values at $S_{\max}=32000$, with substantial statistical uncertainties but a clear, monotonically increasing trend that does not show stabilization even at the largest $S_{\max}$. This behavior demonstrates that the system has not fully saturated its spectral properties and that $\ds$ converges significantly slower than volumetric observables.

By contrast, the Hausdorff dimension $\dhh$ shows a different trend. It increases significantly when moving from shallow ($S_{\max}=4000$) to deeper annealing ($S_{\max}=16000$), but then stabilizes rapidly, showing negligible variation between $16000$ and $32000$ steps. This stability supports the interpretation that the volumetric growth properties of the graphs saturate more quickly than the spectral properties.

The average distance $\mean{d}$ shows a clear trend: at shallow annealing ($S_{\max}=4000$) the graphs are relatively poorly optimized and exhibit larger average distances; as $S_{\max}$ increases to $16000$ and $32000$, $\mean{d}$ decreases and stabilizes, reflecting the combined effect of increased spectral coherence and the edge-density penalty. Overall, these convergence tests indicate that $S_{\max}=16000$ provides a reasonable compromise for volumetric observables, while also highlighting that the spectral dimension has not fully converged in the FSS campaign. Therefore, the reported values of $\ds$ should be interpreted as effective lower bounds on the true ground-state values.

\subsection{Dependence on the curvature coupling $\gamma$ at $N=256$}
\label{subsec:results-gamma}

\begin{figure}[t]
  \centering
  \includegraphics[width=0.7\linewidth]{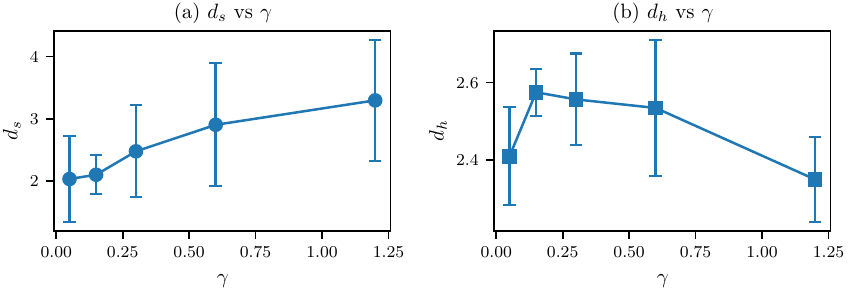}
  \caption{Dependence of the \CCM\ ensemble on the curvature coupling $\gamma$ at fixed size $N=256$, locality coupling $\beta=0.0091$, and annealing depth $S_{\max}=16000$ steps. The panels show the mean energy, algebraic connectivity $\lambda_2$, mean Ollivier--Ricci curvature, spectral dimension $d_s$, Hausdorff dimension $d_h$, and average distance $\mean{d}$ as functions of $\gamma\in\{0.05,0.15,0.30,0.60,1.20\}$. Error bars denote one standard deviation across seeds. Increasing $\gamma$ drives the system towards configurations with higher $\lambda_2$, less negative curvature, and larger $d_s$, while $d_h$ and $\mean{d}$ vary more slowly.}
  \label{fig:GammaSweep}
\end{figure}

We next study the dependence of the emergent geometry on the curvature coupling $\gamma$ at fixed size $N=256$ and fixed locality coupling $\beta=0.0091$. For each value of $\gamma\in\{0.05,0.15,0.30,0.60,1.20\}$ we perform CSA runs with $S_{\max}=16000$ steps and 10 seeds. The aggregated observables are displayed in Fig.~\ref{fig:GammaSweep}.

As $\gamma$ increases, the energy density decreases strongly, indicating that the curvature term plays an increasingly dominant role in driving the system towards low-energy configurations. The algebraic connectivity $\lambda_2$ grows with $\gamma$, reflecting the fact that configurations with less negative curvature tend to also exhibit enhanced global connectivity.

The spectral dimension $\ds$ shows a pronounced increase with $\gamma$, evolving from values around two at the smallest curvature coupling to significantly larger values at the largest couplings. This trend is consistent with the idea that strengthening the curvature term promotes locally more clustered and coherent structures, which in turn support higher effective diffusion dimensions. The Hausdorff dimension $\dhh$, on the other hand, varies more modestly across the range of $\gamma$, remaining in a band around values slightly above two. The average distance $\mean{d}$ exhibits a slow decrease as $\gamma$ increases, in line with the enhanced connectivity and clustering.

Taken together, the $\gamma$-dependence indicates that the curvature coupling is an efficient knob for tuning the spectral properties of the \CCM\ ensemble, while leaving the volumetric dimension and typical distances comparatively more rigid. This separation of roles between curvature and volumetric growth suggests a rich parameter space in which spectral and Hausdorff dimensions can be partially decoupled.

\subsection{Dependence on the locality coupling $\beta$ at $N=256$}
\label{subsec:results-beta}

\begin{figure}[t]
  \centering
  \includegraphics[width=0.7\linewidth]{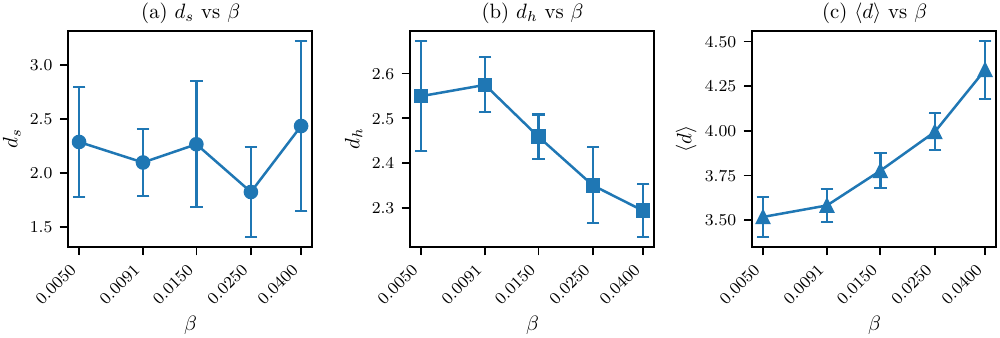}
  \caption{Dependence of the emergent geometry on the locality coupling $\beta$ at fixed size $N=256$, curvature coupling $\gamma=0.15$, and annealing depth $S_{\max}=16000$ steps. Panel (a) shows the spectral dimension $d_s$ versus $\beta$, panel (b) the Hausdorff dimension $d_h$ versus $\beta$, and panel (c) the average distance $\mean{d}$ versus $\beta$, for $\beta\in\{0.0050,0.0091,0.0150,0.0250,0.0400\}$. Error bars denote one standard deviation across seeds. Increasing $\beta$ strengthens the edge-density penalty, leading to sparser graphs with larger typical distances and modest variations in $d_h$, while $d_s$ exhibits a more intricate, non-monotonic response.}
  \label{fig:BetaSweep}
\end{figure}

To probe the role of the edge-density penalty, we perform a sweep in the locality coupling $\beta$ at fixed size $N=256$ and fixed curvature coupling $\gamma=0.15$, with $S_{\max}=16000$ steps and 10 seeds for each value of $\beta\in\{0.0050,0.0091,0.0150,0.0250,0.0400\}$. The resulting dependence of the geometric observables on $\beta$ is shown in Fig.~\ref{fig:BetaSweep}.

The energy density exhibits a clear monotonic trend: more negative values at small $\beta$ gradually increase and eventually become positive as $\beta$ grows. This reflects the competition between the spectral and curvature terms, which favor denser and more coherent graphs, and the edge-count term, which penalizes the addition of edges. As $\beta$ increases, the optimal balance shifts towards sparser configurations with fewer edges.

This is corroborated by the behavior of the algebraic connectivity $\lambda_2$, which tends to decrease with growing $\beta$, indicating that stronger edge penalties reduce global spectral coherence. The Hausdorff dimension $\dhh$ shows a nontrivial but relatively mild dependence on $\beta$, with values that remain in a relatively narrow range around two and a half for the parameter window considered. The average distance $\mean{d}$ increases as $\beta$ grows, illustrating that stronger edge penalties lead to more extended graphs with longer typical path lengths.

The spectral dimension $\ds$ displays a more complex, non-monotonic dependence on $\beta$, with significant statistical uncertainties. At the smallest values of $\beta$ the graphs are relatively dense and spectrally coherent, but may also contain redundant structures that suppress diffusion at intermediate scales. As $\beta$ increases, the system moves towards sparser yet still well-connected configurations, which can either enhance or reduce the effective spectral dimension depending on the detailed interplay between global connectivity and local curvature. The observed pattern suggests that $\beta$ provides a complementary handle, alongside $\gamma$, for shaping the balance between locality, connectivity, and curvature in the \CCM\ ensemble.

\subsection{Consistency checks and distance scaling}
\label{subsec:results-consistency}

\begin{figure}[t]
  \centering
  \includegraphics[width=0.7\linewidth]{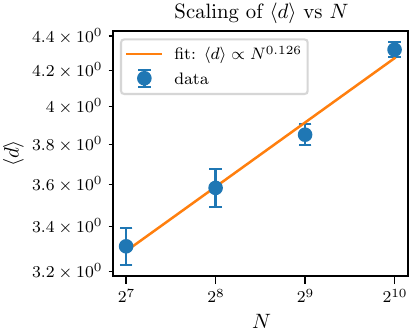}
  \caption{Consistency checks for the \CCM\ ensemble at fixed couplings $(\alpha,\beta,\gamma)=(1.0,0.0091,0.15)$ and annealing depth $S_{\max}=16000$ steps. The panels illustrate the scaling of the average distance $\mean{d}(N)$ across the FSS volumes $N=128$--$\NmaxFSS$, together with a power-law fit of the form $\mean{d}(N)\sim N^\eta$ and the associated residuals. The effective exponent is $\eta=\DistScalingExponent$, indicating a slow growth of typical distances with $N$ consistent with an effectively small-world, yet geometrically nontrivial, regime.}
  \label{fig:Consistency}
\end{figure}

Finally, we perform consistency checks on the scaling of the average distance $\mean{d}(N)$ across the FSS ensembles and extract the effective exponent $\eta$ appearing in Eq.~\eqref{eq:distance-scaling}. The corresponding fits and residuals are summarized in Fig.~\ref{fig:Consistency}.

A fit of $\ln \mean{d}(N)$ versus $\ln N$ over the range $N=128$--$\NmaxFSS$ yields
\begin{equation}
  \eta = \DistScalingExponent.
\end{equation}
This value confirms that typical distances grow slowly with system size, significantly more slowly than any extensive scaling and compatible with an effectively small-world regime over the range of sizes studied. At the same time, the finite and nontrivial Hausdorff dimension and the hierarchy $\ds>\dhh$ indicate that the \CCM\ graphs do not reduce to simple random graphs with purely logarithmic distance scaling, but rather realize a more structured, fractal-like geometry in which spectral and volumetric properties are partially decoupled.

\section{Discussion}
\label{sec:discussion}

The results presented above show that the Coherence--Curvature Model, a hybrid model combining global spectral constraints with local geometric penalties, gives rise to graph ensembles with nontrivial and apparently finite effective dimensions. Several qualitative features emerge robustly across the explored parameter space.

First, at fixed couplings $(\alpha,\beta,\gamma)=(1.0,0.0091,0.15)$, we observe a robust hierarchy between the spectral dimension $\ds$ and the Hausdorff dimension $\dhh$. The Hausdorff dimension $\dhh$ converges rapidly and appears to stabilize around a value close to three. The spectral dimension $\ds$ increases with system size $N$, reaching values that, despite large uncertainties and evidence of incomplete convergence (Sec.~\ref{subsec:results-conv}), consistently exceed $\dhh$. The ratio $d_s/d_h = \DsOverDhNmaxValue \pm \DsOverDhNmaxError$ at $N=\NmaxFSS$ quantifies this hierarchy. This pattern is reminiscent of scenarios in discrete quantum gravity and random-graph models where the spectral dimension exceeds the Hausdorff dimension, signaling the presence of nontrivial connectivity and effective shortcuts that enhance diffusion~\cite{Ambjorn:2012CDT,Bollobas:2001RandomGraphs,Barabasi:2016book}.

Second, a key finding is the significant decoupling between spectral and volumetric dimensions revealed by the dependence on the curvature coupling $\gamma$. The analysis in Sec.~\ref{subsec:results-gamma} demonstrates that the curvature term is an efficient control parameter for the spectral properties of the ensemble. Increasing $\gamma$ drives the system towards configurations with higher algebraic connectivity and less negative Ollivier--Ricci curvature, and it correlates with a significant growth of $\ds$, while leaving $\dhh$ and $\mean{d}$ comparatively stable. This demonstrates that the CCM provides a mechanism (via local curvature) to tune the global diffusion properties almost independently of the volumetric growth.

Third, the locality coupling $\beta$ plays a distinct but complementary role. By directly penalizing the number of edges, $\beta$ controls the density and locality of the graphs. The BETA sweep (Sec.~\ref{subsec:results-beta}) shows that increasing $\beta$ leads to sparser configurations with larger average distances and reduced algebraic connectivity. The Hausdorff dimension responds gently, while the spectral dimension exhibits a complex dependence on $\beta$.

Fourth, the scaling of the average distance with system size, characterized by the small effective exponent $\eta=\DistScalingExponent$, points to a regime in which typical path lengths grow very slowly with $N$. This behavior is strongly indicative of small-world properties. The observed hierarchy $d_s > d_h$ is characteristic of such regimes, where long-range connections (shortcuts) enhance diffusion efficiency relative to the underlying volumetric structure. The CCM Hamiltonian inherently generates this geometry by balancing local sparsity (driven by $\beta$ and $\gamma$) against the explicit demand for global coherence (driven by $\alpha \lambda_2$).

From the perspective of emergent geometry and quantum gravity, these findings illustrate how the competition in the \CCM\ Hamiltonian can generate rich geometric behavior. The presence of a robust $d_h \approx 3$ combined with a tunable $d_s > d_h$ suggests that \CCM-like models may realize universes with scale-dependent dimensionality, qualitatively similar to phenomena observed in other discrete approaches~\cite{Ambjorn:2012CDT,Oriti:2011GFT}.

At the same time, the convergence studies highlight the challenges inherent in extracting reliable effective dimensions. The spectral dimension in particular displays slow convergence and large uncertainties, underscoring that the reported values are likely lower bounds. Furthermore, the maximum system size ($N=\NmaxFSS$) limits the separation of scales, with the average diameter remaining very small (slightly above four at $N=\NmaxFSS$). This lack of a clear intermediate regime means the extracted dimensions are still influenced by finite-volume and short-scale effects. Nevertheless, the qualitative trends reported here, particularly the decoupling of $d_s$ and $d_h$, are robust, providing a coherent picture of the \CCM\ as a numerical laboratory for studying geometrogenesis.

\section{Conclusion}
\label{sec:conclusion}

We have introduced and analyzed the Coherence--Curvature Model, a dynamical ensemble of simple, undirected, connected graphs governed by a Hamiltonian that combines a spectral coherence term $-\alpha\,\lambda_2(\Graph)$, an edge-count penalty $+\beta\,\abs{\Edges}$, and a curvature term $\gamma\sum_v \Ric(v)$ based on Ollivier--Ricci curvature. Using connected simulated annealing, we have generated low-energy configurations for a range of system sizes and coupling values, and we have characterized their emergent geometry through the spectral dimension $\ds$, the Hausdorff dimension $\dhh$, and the average graph distance $\mean{d}$.

At the largest simulated volume $N=\NmaxFSS$ and for the reference couplings $(\alpha,\beta,\gamma)=(1.0,0.0091,0.15)$, we find
\[
  d_s = \DsNmaxValue \pm \DsNmaxError,\qquad
  d_h = \DhNmaxValue \pm \DhNmaxError,\qquad
  \frac{d_s}{d_h} = \DsOverDhNmaxValue \pm \DsOverDhNmaxError.
\]
These estimates are compatible, within current uncertainties, with an effective spectral dimension approaching four, while the Hausdorff dimension remains significantly lower but of the same order, indicating a clear hierarchy between spectral and volumetric dimensions in the \CCM\ ensemble. The average distance $\mean{d}(N)$ exhibits a slow power-law growth with effective exponent $\eta=\DistScalingExponent$, consistent with an effectively small-world regime that nonetheless retains a finite volumetric dimension.

A central result is the observed decoupling between spectral and volumetric dimensions. By varying the curvature coupling $\gamma$, we have demonstrated the ability to tune the spectral dimension $\ds$ largely independently of the Hausdorff dimension $\dhh$. This highlights a novel mechanism where local geometric constraints (curvature) can control global diffusion properties. The locality coupling $\beta$ provides an additional control parameter that modulates edge density and locality. Together, the $(\gamma,\beta)$ plane defines a rich phase diagram in which spectral and volumetric properties can be tuned independently.

The present study has several limitations. The accessible system sizes are modest, and the connected-simulated-annealing dynamics converges slowly, especially for spectral observables. The use of Ollivier--Ricci curvature with a fixed idleness parameter and entropic-regularized optimal transport introduces additional algorithmic choices that may influence quantitative results. Nonetheless, the qualitative picture of a finite-dimensional, fractal-like emergent geometry with a robust $d_s>d_h$ hierarchy appears stable across these limitations.

Future work will aim at extending the simulations to larger volumes and deeper annealing schedules, exploring alternative curvature prescriptions and idleness parameters, and systematically mapping the $(\gamma,\beta)$ phase diagram. It would also be interesting to investigate scale-dependent effective dimensions (e.g., via diffusion-time dependent $\ds(t)$), to couple matter or additional degrees of freedom to the \CCM\ graphs, and to compare the resulting geometries more directly with those arising in causal dynamical triangulations, group field theories, and other discrete approaches to quantum gravity~\cite{Ambjorn:2012CDT,Oriti:2011GFT}.

\section*{Acknowledgments}

The author acknowledges the use of open-source Python libraries, including \textsc{NetworkX}, \textsc{GraphRicciCurvature}, and \textsc{POT}, which made the simulations and analysis in this work possible. No specific external funding was received for this research.

\section*{Data Availability Statement}

The source code used for simulations and analysis, along with the aggregated data supporting the findings of this study, will be made available in a public Zenodo repository; the final URL and DOI will be provided upon publication.

\bibliographystyle{apsrev4-2}
\bibliography{bib/references}

@incollection{Ambjorn:2012CDT,
  author    = {J. Ambj{\o}rn and J. Jurkiewicz and R. Loll},
  title     = {Causal Dynamical Triangulations and the Quest for Quantum Gravity},
  booktitle = {Foundations of Space and Time: Reflections on Quantum Gravity},
  editor    = {G. Ellis and J. Murugan and A. Weltman},
  publisher = {Cambridge University Press},
  year      = {2012},
  pages     = {321--337},
  doi       = {10.1017/CBO9780511920998.013}
}

@incollection{Oriti:2011GFT,
  author    = {Daniele Oriti},
  title     = {The Microscopic Dynamics of Quantum Space as a Group Field Theory},
  booktitle = {Foundations of Space and Time: Reflections on Quantum Gravity},
  editor    = {G. Ellis and J. Murugan and A. Weltman},
  publisher = {Cambridge University Press},
  year      = {2012},
  pages     = {257--320},
  note      = {See also arXiv:1110.5606}
}

@book{Bollobas:2001RandomGraphs,
  author    = {B{\'e}la Bollob{\'a}s},
  title     = {Random Graphs},
  series    = {Cambridge Studies in Advanced Mathematics},
  volume    = {73},
  edition   = {2},
  publisher = {Cambridge University Press},
  address   = {Cambridge},
  year      = {2001},
  isbn      = {978-0521797221}
}

@book{Barabasi:2016book,
  author    = {Albert-L{\'a}szl{\'o} Barab{\'a}si and M{\'a}rton P{\'o}sfai},
  title     = {Network Science},
  publisher = {Cambridge University Press},
  address   = {Cambridge},
  year      = {2016},
  isbn      = {978-1107076266}
}

@article{Ollivier:2009Ricci,
  author    = {Yann Ollivier},
  title     = {Ricci Curvature of Markov Chains on Metric Spaces},
  journal   = {Journal of Functional Analysis},
  volume    = {256},
  number    = {3},
  pages     = {810--864},
  year      = {2009},
  doi       = {10.1016/j.jfa.2008.11.001}
}

@article{Ni:2019Ricci,
  author    = {Chien-Chun Ni and Yu-Yao Lin and Jie Gao and Xianfeng David Gu},
  title     = {Community Detection on Networks with {R}icci Flow},
  journal   = {Scientific Reports},
  volume    = {9},
  number    = {1},
  pages     = {9984},
  year      = {2019},
  doi       = {10.1038/s41598-019-46380-9}
}

@inproceedings{Cuturi:2013Sinkhorn,
  author    = {Marco Cuturi},
  title     = {Sinkhorn Distances: Lightspeed Computation of Optimal Transport},
  booktitle = {Advances in Neural Information Processing Systems 26 (NeurIPS 2013)},
  editor    = {C. J. C. Burges and L. Bottou and M. Welling and Z. Ghahramani and K. Q. Weinberger},
  pages     = {2292--2300},
  year      = {2013},
  url       = {https://proceedings.neurips.cc/paper/2013/hash/af21d0c97db2e27e13572cbf59eb343d-Abstract.html}
}

@article{Flamary:2021POT,
  author    = {R{\'e}mi Flamary and Nicolas Courty and Alexandre Gramfort and Mokhtar Z. Alaya and Aur{\'e}lien Boisbunon and Stanislas Chambon and Laetitia Chapel and Adrien Corenflos and Kilian Fatras and Nemo Fournier and others},
  title     = {POT: {P}ython Optimal Transport},
  journal   = {Journal of Machine Learning Research},
  volume    = {22},
  number    = {78},
  pages     = {1--8},
  year      = {2021},
  url       = {http://jmlr.org/papers/v22/20-451.html}
}

\end{document}